\documentclass[11pt,a4paper]{JHEP3}

\usepackage{amssymb,amsmath}
\usepackage{epsfig}
\usepackage{cite}
\usepackage{listings}

\def\CN{{\cal N}}

\def\tr{{\rm tr}}

\def\a{\alpha}\def\b{\beta}\def\g{\gamma}
\def\d{\delta}\def\e{\epsilon}

\def\l{\lambda}

\def\G{\Gamma}

\newcommand{\vev}[1]{\left\langle{#1}\right\rangle}

\newcommand{\be}{\begin{eqnarray}}
\newcommand{\ee}{\end{eqnarray}}
\newcommand{\nn}{\nonumber}

\title{ 1/4 BPS String Junctions and \boldmath $N^3$ Problem \\
in  6-dim (2,0) Superconformal Theories}
\author{
Stefano Bolognesi$^a$ and Kimyeong Lee$^b$ \\
Department of Applied Mathematics and Theoretical Physics, \\
University of Cambridge, UK$^a$\\ Korea Institute for Advanced Study,\\ Seoul 130-722, Korea$^b$\\
{\tt s.bolognesi, klee@kias.re.kr}
}

\abstract{ We explore 1/4 BPS objects in the Coulomb phase of the ADE-type 6-dim (2,0) superconformal theories. By using the previous work on the junctions of  strings in 5-dim gauge theories and 6-dim superconformal theories, we count the number of    1/4 BPS objects, which     are made of waves on selfdual strings and junctions of selfdual strings and show that for all cases the number   matches exactly one third of the anomaly constant $c_G=d_G h_G$  which is the product of dimension $d_G$ and dual Coxeter number $h_G$.    This suggests the long sought after $N^3$ degrees of freedom are these 1/4 BPS objects  at least in the Coulomb phase.}

\begin{document}

\section{Introduction}

The 6-dim (2,0) superconformal field theories are an important cornerstone for the M-theory structure and for the whole hierarchy of supersymmetric field theories. They come as the ADE type and are realized as the low energy dynamics of type IIB string theory on ADE type singularities \cite{Witten:1995zh}. The AD type theories are also realized  as the low energy dynamics on parallel M5 branes, maybe with OM5 orientifold. These theories are purely quantum and the exact nature of the nonabelian description is not known. It is known from gravity dual that the entropy scales like $N^3$ for $A_{N-1}=SU(N)$   theory \cite{Klebanov:1996un,Henningson:1998gx}. In the generic Coulomb phase, there exist obviously 1/2 BPS selfdual strings of order $N^2$ kinds.  Sometime ago it has been suggested that this additional degrees of freedom could manifest in the Coulomb phase as  1/4 BPS three monopole-string junctions~\cite{Lee:2006gqa}. We want here to make further developments and elaborations on this idea.

A finer counting of the degrees of freedom was proposed  by the anomaly calculation under the global $SO(5)$ R-symmetry of the (2,0) theories \citation{Harvey:1998bx,Yi:2001bz,Intriligator:2000eq}. This has been further supported by recent works on M5 branes via the conformal field theory~\cite{Bonelli:2009zp,Alday:2009qq}. The anomaly coefficient for the ADE type is the product $c_G=d_G*h_G$ of  dimension  $d_G$ and   dual Coxeter number $h_G$. The number $c_G$ for the $ADE$ type group is divisible by $6$. In this work, we explore the 1/4 BPS objects  in the generic Coulomb phase of the (2,0) theories of ADE type. Irreducible 1/4 BPS objects consist of    BPS waves on selfdual strings and junctions made of three selfdual  strings. While  a selfdual string can turn to its anti-selfdual string by a spatial rotation, these 1/4 BPS objects and its anti-objects are distinguishable. In this work we count  the number of 1/4 BPS objects and its anti-objects   by considering their charges only, and found the number is exactly $c_G/3$ for all ADE types theories.   Our counting suggests that there may be no further degrees of freedom to account.   

In Ref~\cite{Lee:2006gqa}, it has been argued that there are further less supersymmetric nonplanar BPS webs of selfdual strings. Their basic elements are 1/4 BPS junctions and so they may be not included as the fundamental degrees of freedom.  The key for these BPS webs of strings is the locking of the internal $SO(5)$ R-symmetry with the spatial $SO(5)$ rotational symmetry.
  In Ref.~\cite{Intriligator:2000eq}, the $SO(5)_R$ anomaly is studied in the Coulomb phase by the Wess-Zumino-Witten term for the five scalar fields, where selfdual string appears as skyrmions with the topology $\pi_4(S^4)={\mathbb Z}$.   Our match  of two numbers suggests that there may be a way to include 1/4 BPS junctions to this argument. (See~\cite{Boyarsky:2002ck} for the anomaly analysis for the monopole strings in 5-dim gauge theory.)  There are somewhat different approach the Coulomb phase. One could consider the Wilsonian effective Lagrangian for the abelian modes, which would be expressed in terms of the derivative expansion. The contributions to higher order derivative terms could contain various bubbles of BPS and anti-BPS objects. It would be great if one   sorts out  the leading nontrivial terms.

Another approach to the (2,0) theories is to compactify them on a circle of radius $R_6$. Its low energy Lagrangian is 5-dim maximally supersymmetric gauge theories.  The dimensionful coupling constant is order of $R_6$ so that Kaluza-Klein modes of the underlying (2,0) theory manifest as instanton solitons in 4+1 dimensions~\cite{Seiberg:1996bd}.  Especially for the A-type low energy theory on $N$ M5 branes become that of $N$ D4 branes, which is the $SU(N)$ gauge theory. 
 This 5-dim $SU(N)$  theory is weakly coupled in the low energy, superficially divergent in ultraviolet, and so UV incomplete.  Recently there has been a proposal that the 5-dim theory could in fact be much more complete than what superficially thought\cite{Douglas:2010iu,Lambert:2010iw}.   
 Also the 5-dim theory contains selfdual strings as monopole strings. If this claim true, then it certainly means that the $N^3$ degrees of freedom of the 6-dim theory must  already be present in the 5-dim theory. Since the 5-dim gauge theory entropy scales like $N^2$, at least at small energy, the $N^3$ states must be hidden in some way.
Our paper is in part stimulated by these recent conjectures although we will not rely directly on them in our arguments.

So what could be the possible solutions to the $N^3$ problem? We can spit the possible solutions in two  categories:  {\rm (I)} The fundamental degrees of freedom are not present in the 5d description at all. They must thus be in the extra KK modes and not accessible when the temperature is small.  {\rm (II)} The fundamental degrees of freedom are already there but hidden in some way.

There are some reasons to be skeptical about option {\rm (I)}. It would seem strange that the $N^3$ degrees of freedom are not present at all in the KK zero modes. In this case we would be forced to consider two different kind of fundamental degrees of freedom, t  the ones of the 5d theory and the $N^3$ ones in higher KK modes. But  instantons of the 5d theory are KK modes.
There may be more KK modes besides those captured by instantons, but there is no concrete evidence for that yet. The naive counting of BPS instantons does not generate $N^3$. Classically they have many zero modes but quantum mechanically, one expect that there is only single threshold bound state for each instanton number besides the spin counting. There is a proposal that instantons are made of $N$ instanton partons of mass $1/(NR_6)$ and instanton partons fall to the adjoint representation~\cite{Collie:2009iz,Tong:2010zz}. The number of instanton partons would then lead  to the $N^3$ counting. Recently we have proposed more concrete realization of this instanton partons~\cite{Bolognesi:2011kl},  but there is no evidence yet that instanton partons are in adjoint representation. Especially the instanton partons are intrinsically related to the compactification as they are presumably arising from a single M5 brane wrapping the compactification circle $N$ times. While there may be a room for instanton partons to play in the 5-dim gauge theories, it is not clear yet what role they would play in 6-dim. (See somewhat completely different approach to introduce $N^3$ d.o.f. in 3,4 dim~\cite{Leigh:2010va}. Also for somewhat interesting approach to nonabelian tensor field, see Ref.~\cite{Ho:2011sd}.)

Option {\rm (II)} sounds better but still requires some further specifications. Additional states could be formed as a bound or confinement of BPS objects, for example, dyonic instantons. When temperature goes up, the bound states would be broken to their fundamental components. However the $N^3$ entropy is what we see even at arbitrary   high temperature.  Furthermore we should also have a good reason to select $N^3$ instead of $N^4$ or $N^5$. Bounded states can be created with any number of legs and so there seems to be no criterion to prefer $N^3$. 

We thus conclude   that $N^3$ states if they emerge from somewhere must be strange, at least at first glance. First they must have three legs, and being absolutely stable, no matter what the energy  is. Higher leg object could exists, but they must be just composite of the three leg ones. Second they cannot be states with finite energy. This could sound is contrast to the fact that the entropy count the number of states {\it up} to a certain energy, and infinite energy objects are not counted.    
 
One way out is that these states could be  confined objects like quarks or gluons in QCD. The transition from $N^2$ to $N^3$ of the entropy as the temperature or the energy involved is increase in 5-dim gauge theories, could be analogue  to the deconfinement transition in QCD where the entropy jumps from $N^0$ to $N^2$ 4-dim gauge theories. The energy, or free energy, of a single quark is divergent in the confined phase. 
Is enough to think about the electric flux, being confide is a flux-string, gives linearly divergent energy. The free energy of a single quark becomes instead infinite when the temperature reaches the deconfinement transition and so they can count in the entropy. Actually the gluons dominate in the count because they are of order $N^2$.

Infinite energy object with three legs have in fact been found in \cite{Lee:2006gqa} as 1/4 BPS  junctions of monopole strings in the Coulomb phase of  the 5-dim theories.   The junctions has all the requirements we are looking for to be a good candidate for the $N^3$ degrees of freedom. They are already present  in the 5d theory. It has three legs, and BPS objects with more legs being  just composite objects of the junctions. It has infinite energy due to the string legs. Also there are 1/4 BPS waves on monopole strings. 
 
 Still many question remain to be answered though. What is the exact nature of this deconfinment phase transition? The flat direction disappears in any finite temperature, leading to the gauge symmetry restoration.  At  zero temperature, junctions and anti-junctions would be bounded by the linear potential and so confined.  Interesting question which we do not attempt here is to count the number of independent unstable massive mesons in the Coulomb phase of 5d theories. or in 6d theories. They  would contribute to   the scattering amplitude and the high derivative   low energy effective Lagrangian for the abelian degrees of freedom.

 The anomaly polynomial under the general background of the $SO(5)_R$ gauge field strength $F$ and graviational curvature $R$    for a single M5 brane\cite{Witten:1996hc} is
 \be I_8(1)= \frac{1}{48}\Big[p_2(F)-p_2(R)+\frac14(p_1(F)-p_1(R))^2\Big],
 \ee
where $p_k$ is the k-th Pontryagin class.  The similar anomaly polynormial for the 6-dim (2,0) theories of the group $G$   is calculated for AD-type and also conjectured for the $E$-type  to be~\cite{Harvey:1998bx,Intriligator:2000eq,Yi:2001bz} 
\be I_8[G]= r_G I_8(1)+     c_G \times \frac{p_2(F)}{24}, \ee
where $r_G$ is the rank of the group and $c_G$ is the anomaly coefficient. 
 The anomaly coefficient $c_G$ is  the product of the dimension $d_G$ of the
group and the dual Coxeter number $h_G$
\be c_G= d_G * h_G.
\ee
Table I  enlists all $r_G,d_G, h_G$ and $c_G/3$ for the ADE type groups.
\begin{center}
\begin{tabular}{|c|ccc|c|}
\hline
Group &  $r_G$ &   $d_G$ & $h_G$ &     $c_G /3 $  \\
\hline
$A_{N-1}=SU(N)$ &  $ N-1$ &  $N^2-1$ & $N$ &  $\frac13 N(N^2-1)$ \\
$D_N=SO(2N)$ & $ N$ &  $N(2N-1)$ & $2(N-1)$ &   $\frac23 N(2N-1)(N-1)$ \\
$E_6$ &  6 &  78 & 12 &   312 \\
$E_7$ & 7 &   133 & 18 &    798  \\
$E_8$ &  8 & 248 &  30      & 2480  \\
\hline
\end{tabular}
\vskip 1cm
\centerline{Table I: $r_G$, $d_G$, $h_G$ and $c_G/3$ for simple-laced groups $ADE$}
\end{center}

More recently there were several works exploring the M5 branes wrapped on a certain kind   of  4-dim manfolds, resulting in 2d (2,0) superconformal field theory with $U(1)$ R-symmetry. The central charge of the Toda theory of similar gauge group  takes a  similar structure~\cite{Bonelli:2009zp,Alday:2009qq} as it is given as
\be c_{\rm Toda}[G]= r_G+ c_G Q^2, \label{toda} \ee
where $Q=(\e_1+\e_2)^2/(\e_1\e_2)$ if the 6-dim theory is compactified on   $R^4$ with equivariant parameters $\e_{1,2}$. 

The plan of the paper is as follows. In Sec.2, we first review the BPS junctions in 5-dim gauge theories. In Sec.3,  we count 1/4 BPS objects and anti-objects in the (2,0) theories of ADE types. Finally in Sec.4, we close with some concluding remarks.

\section{5-dim Gauge Theories and Junctions}

We first consider the 1/2 BPS and 1/4 BPS objects in 5-dim gauge theories as we have their Lagrangian. Then we consider the strong coupling limit of these BPS objects as identify them as
BPS objects in the 6-dim (2,0) theories. Then, we count 1/2 BPS and 1/4 BPS objects in the 6-dim (2,0) theories for ADE types and compare these numbers with the anomaly coefficient.

Let us start with   the $(0,2)$ $A_{N-1}=SU(N)$ superconformal theory in $5+1$ which rises from the low energy of $N$ M5 branes in M-theory. Upon compactification to a circle with radius $R_5$ it reduces to the low-energy of $N$ D4 branes in type IIA. This is maximal symmetry $SU(N)$ Yang-Mills theory in $4+1$ with coupling constant 
\be
8\pi /g_{5}^2 = 1/R_{6}.
\ee
The theory has sixteen supercharges and consist of a gauge multiplet $A_{\mu}$ together with five scalars $\phi_I$ with $I=6,7,8,9,10 $ and fermion superpartners $\Psi$. The bosonic part of Lagrangian is 
\be
L_B= \frac{1}{2 g_5^2}   \tr(F_{MN}F^{MN}  +2 D_M\phi_I D^M\phi_I + [\phi_I,\phi_J]^2  ).
\ee
We have three main topological charges in the Coulomb phase: the electric charge $Q_E$, the magnetic charge $Q_M$ of monopole strings and the instanton charge $Q_{{\cal I}}$. The supersymmetry transformation for the gluino is
\be \d\l = \frac12 \G^{MN}F_{MN} + \G^{MI}D_M\phi_I -  \frac{i}{2} \G^{IJ} [\phi_I,\phi_J]\ee
in 10-d supersymmetry notation. Under the $SO(5)$ R-symmetry, the spinors transform as ${\bf 4}$ and $\bar{\bf 4}$ and the scalars transform as ${\bf 5}$. When the scalar fields take mutually commuting nonzero expectation value, the flavor symmetry and gauge symmetries are spontaneously broken. The diagonalized scalar vacuum expectation values denote the position of M5 or D4 branes in normal $R^5$ space modulo dimensionful parameter. The obvious BPS objects are W-bosons, monopole strings, and instantons.

In the $SU(2)$ gauge theory on two separated D4 branes with $\vev{\phi_1}=(v, -v)/2$, there exist    1/2 BPS monopole strings, corresponding to D2 branes connecting two   D4 branes. For a single string along $x^4$ axis would be described by the selfdual equation
\be F_{ij} = \e_{ijk}D_k \phi_1, \ \ i,j,k\in \{1,2,3\} , \ee
and the unbroken susy parameter satisfies
\be \G^{1235}\e= \e . \ee
Its tension would be  
\be T_{12} = \frac{4\pi v}{g_5^2} .   \ee
Interestingly the anti-monopole strings are related to the monopole strings as one can rotate $x^3,x^4$ axis by 180 degrees, which changes the sign flip of the supersymmetry. This  is related to the fact that  the monopole strings can  make a closed loop, which has the same quantum number as vacuum. In the strong coupling limit, the monopole strings would become 1/2 BPS selfdual strings once we keep $v/g_5^2$ constant, which is related to the relative position between two M5 branes.

There exist  massless particles form 1/2 BPS short multiplet. Also there would be 1/2 BPS W-bosons and 1/2 BPS instantons whose conserved susy parameteres are, respectively,
\be \G^{05}\e=\e, \ \ \G^{1234}\e=\e  .\label{dyonicinstanton}\ee
The 5-dim W-bosons in the Coulomb phase can be intepreted as    self-dual strings wrapping the compactified circle, and The 5-dim instanton solitons  are just KK modes along $x^5$ direction. 
One could also think about the 1/2 BPS dyonic strings which are monopole strings with uniform electric charge density. These 1/2 BPS dyonic strings describe  the selfdual strings  partially wrapping $x^{5}$ direction, and so their direction is tilted in $x^4,x^5$ plane.

Now let us consider the 1/4 BPS objects in the 5-dim $SU(2)$ gauge theory. The 1/4 BPS states can be dyonic instantons~\cite{Lambert:1999ua,Kim:2006ee} which is invariant under parameters satisfying both conditions in Eq.~\ref{dyonicinstanton}.  Similarly, one could have 1/4 BPS  left or right moving waves on  magnetic monopoles whose supersymmetric parameter  satisfies   
\be \G^{1235}\e=\e, \ \G^{04}\e=\e . \ee
A way to approach the wave on the monopole string is to consider the zero modes of the BPS monopoles, and lift them to the zero modes of monopole strings. Instead of the position and phase of BPS monopoles, we would get the massless modes along monopole strings. Both dyonic instantons and waves on monopole strings correspond to the 1/4 BPS left or right moving waves on selfdual strings in the 6-dim theory. The selfdual strings with left (right)   moving wave  form 1/4 BPS (anti-)object.   The anti 1/4 BPS object would satisfy the   supersymmetric condition   $\G^{1235}\e=\e, \G^{04}\e=-\e$ with one of the sign got flip.  There is no 1/2 BPS junctions as there exists only one kind of selfdual string in the $SU(2)$ case. Thus for $A_1=SU(2)$, the number of 1/4 BPS object in the 6-dim theory is $2=C_{A_1}/3$.

Let us first consider the 5-dim theory with $A_{N-1}=SU(N)$ gauge group. There exist  1/2 BPS monopole strings for any root, say, $\a=e_i-e_j$ in the generic Coulomb phase.   It would correspond to the D2 branes connecting $i$-th D4 brane to $j$-th D4 branes. As none of three D4 branes are lined up, there would be only four zero modes for such 1/2 BPS monopole strings. The conserved supersymmetries  of 1/2 BPS strings would be different from each other as they are not parallel. The tension of the 1/2 BPS monopole strings would be proportional to the distance between two D4 branes. Going to the 6d (2,0) theories in the  generic Coulomb phase, we expect that there are 1/2 BPS selfdual strings with four zero modes and tension  
\be T_\a \sim |\a \cdot \phi_I| . \ee
The string for the opposite $-\a$ root is again obtained just by a spatial rotation.

\begin{figure}[h]
\begin{center}
\leavevmode
\epsfxsize 10 cm
\epsffile{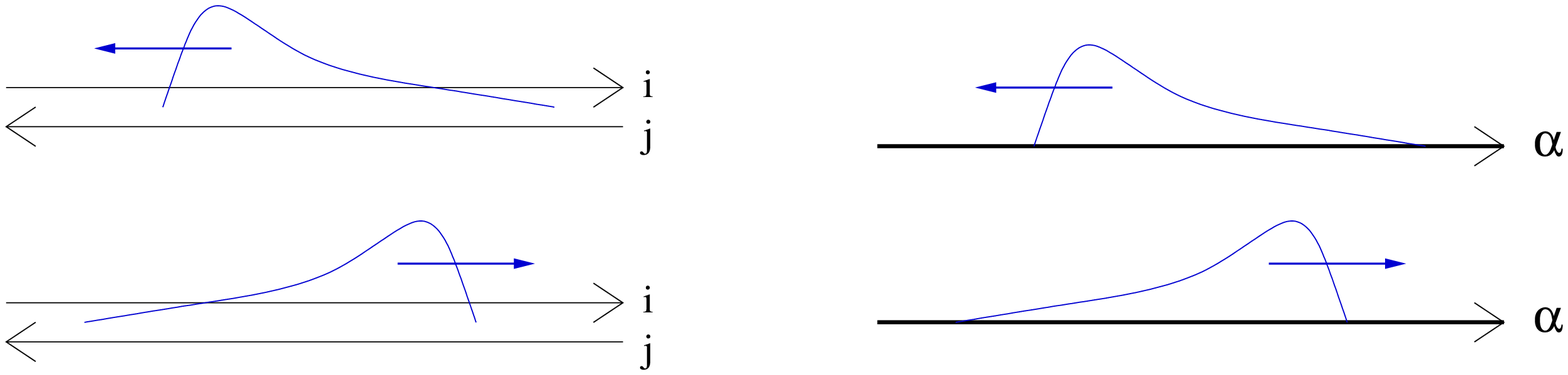}
\end{center}
\caption{\footnotesize two representations of 1/4 left and right moving BPS waves on a selfdual string}
\label{chiralwave}
\end{figure}

 From our understanding of the left and right moving waves on  monopole strings and also dyonic instantons, one see the 1/4 BPS left or right moving waves on selfdual strings could have a finite transverse energy profile. As the wave is moving with speed of light the profile of the wave is stationary and has no dissipation.   Fig.\ref{chiralwave} shows two kinds of representations for both left and right moving waves on the selfdual string corresponding to the root $\a= e_i-e_j$. 
%
 
The minimally supersymmetric, or 1/16, BPS webs of selfdual strings arise from the locking of the spatial $SO(5)$ rotation to SO(5) R-symmetry. In 5d Yang-Mills theory, the BPS equation for the BPS dyonic webs of  monopole strings~\cite{Lee:2006gqa} is
\be && F_{ab}-\e_{abcd}D_c\phi_d + i[\phi_a,\phi_b]=0, \ D_a\phi_a=0 ,  \nn \\
&& F_{a0}=D_a\phi_5, \ \ D_a^2\phi_5 -[\phi_a,[\phi_a,\phi_5]]=0 ,
\ee
where $a,b,c,d=1,2,3,4$ and the Gauss law is used with the gauge $A_0=\phi_5$. Here we renamed scalar field with shift of indices $\phi_I \rightarrow \phi_{I-5}$. 

\begin{figure}[h]
\begin{center}
\leavevmode
\epsfxsize 12 cm
\epsffile{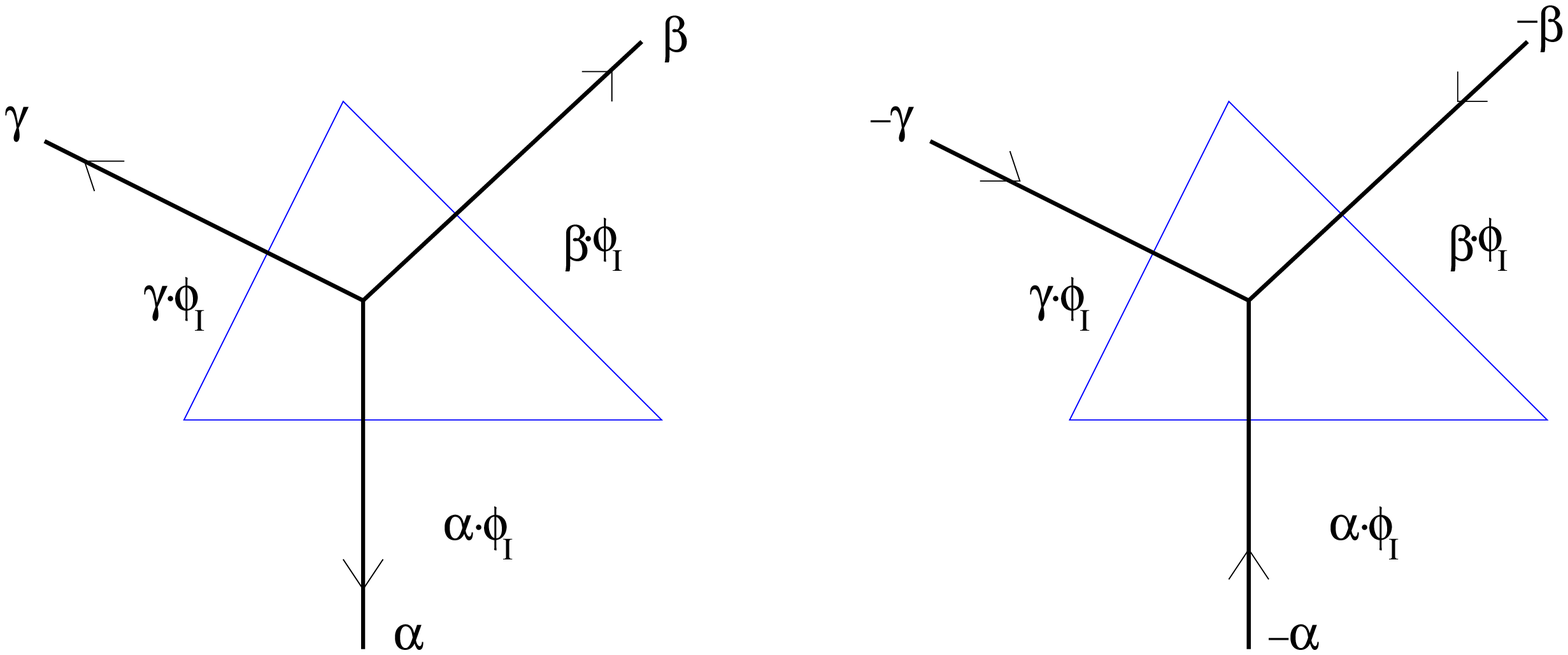}
\end{center}
\caption{\footnotesize Triangle and 1/4 BPS Junction and Anti-Junction}
\label{abcjunction}
\end{figure}

Simplest ones are of course 1/4 BPS planar junctions of monopole strings~\cite{Lee:2006gqa}, which can exist when $ N\ge 3$ for the   $SU(N)$ group.   One needs generic   scalar expectation value so that any given  three M5 branes characterized by indices $i,j,k$ are not aligned. There would be three corresponding roots $\a=e_1-e_2,\b=e_2-e_3,\g=e_3-e_1$ such that sum of these three roots vanish. There would be    1/2 BPS selfdual strings for each roots. Once three strings are on the plane and form a junction such that the junction form a dual lattice to the triangle defined by $\a\cdot\phi_I,\b\cdot\phi_I,\g\cdot\phi_I$, the tension of selfdual string gets balanced and the junction becomes 1/4 BPS~\cite{Lee:2006gqa}. Figure~\ref{abcjunction} shows both such 1/4 BPS junctions and anti-junctions.
 Anti-junction has the opposite 
charge orientation. Dyonic string webs would be also 1/4 BPS as they can be obtained from the selfdual string webs tilted along $x^5$ direction and stacked periodically along $x^5$ direction.%
%

\begin{figure}[h]
\begin{center}
\leavevmode
\epsfxsize 10 cm
\epsffile{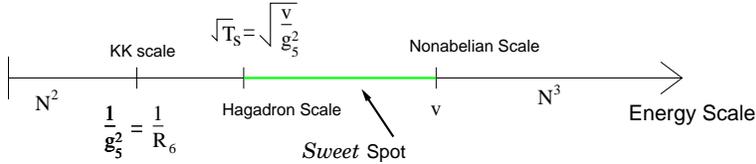}
\end{center}
\caption{\footnotesize Energy scale in the Coulomb phase}
\label{energyscales}
\end{figure}

Before we count the 1/4 BPS objects in the (2,0) theories, let us consider an  important issue of the energy scales of the theory in the Coulomb branch of 5d theories. What follows is     summarized in Fig \ref{energyscales}.
The Coulomb branch becomes quantitative tractable when the brane separation is big enough so that
\be
v \gg 1/g_5^2 .
\label{regime}
\ee
At low energies the theory is 5d abelian gauge theory. At the KK scale $1/R_6\sim 1/g_5^2$ the theory becomes 6-dim but still Abelian. the monopole-strong become the massive self dual string but still is heavy enough not to affect the counting of degrees of freedom. The next stage is at the Hagedorn scale given by $\sqrt{T_s}$ with string tension $T_s\sim \sqrt{v/g_5^2}$.  At this scale string and anti-string get deconfined. Similary one expects junctions and anti-junctions get deconfined. Note that precisely because we are working in the regime (\ref{regime}), the Hagedorn scale is still much smaller than the inverse thickness of the string $v$. Thus we can consider  strings and junctions as fundamental and the Hagedorn phase transition follows. The next stage is the scale $v$ or $W$ boson mass  where the theory becomes non-abelian 6-dim and the strongly coupled. The known fact  is that the entropy is $N^3$ at this scale. Our observation is that the jump to $N^3$ happens actually before  at the Hagedorn scale. The region from $\sqrt{T_s}$ to $v$ is our {\it  sweet} spot, here the theory is still weakly coupled but nevertheless the $N^3$ degrees of freedom are active in the entropy computation. 

Another  way to approach the nonabelian degrees of freedom is to find the 5d low energy effective Lagrangian for the abelian degrees of freedom. While there is the usual perturbative contributions by W-boson loops, one expects also the nonperturbative contributions by instanton loops, selfdual string virtual bubbles and also by junction bubbles. The ultraviolet completeness of the 5d theories, if it is true, should include all these nonperturbative effects.

\section{Counting $\frac14$ BPS Objects in ADE Type (2,0) Theories}   

We now start to  count the number of 1/4 BPS objects. As we count the infinite energy objects, they
are not states with some spin structures. Rather they are characterized by what kinds of charge they carry. 
Let us recall how to count the degrees of freedom for the 4d $SU(N)$ gauge theory on $N$ D3 branes in the Coulomb phase in the weak coupling limt. First of all   the $SU(N) gauge$ symmetry is spontaneously broken in the Coulomb phase to $U(1)^{N-1}$, and we find the matters in the adjoint representations, leading to the $N^2-1$ distinct states:  for each pair of distinct branes we have a charged W boson and anti W-boson, which count as $N(N-1) $. For    each brane we have a photon which count as $N-1$, subtracting the global $U(1)$. Total light 1/2 BPS object is $N-1+ N(N-1)= N^2-1$, corresponding to the adjoint representation.  

 For the (2,0) theory in the generic Coulomb phase with gauge group $A_{N-1}=SU(N)$. The $N(N-1)$ root vectors of the Lie algebra $A_{N-1}$ can be represented by $e_i-e_j$, where $e_i$ with $i=1,\cdots N$ are  $N$-dim orthonormal vectors. 
There are 1/2 BPS $N-1$ massless particles or waves corresponding to the Cartan elements of the Lie algebra and there are 1/2 BPS $N(N-1)/2$ selfdual strings. 1/2 BPS massless particles of $N-1$ kinds have finite energy, but 1/2 BPS strings of $N(N-1)/2$ kinds would have infinite energy. It is hard to say they form an adjoint representation.
 
\begin{figure}[h]
\begin{center}
\leavevmode
\epsfxsize 6 cm
\epsffile{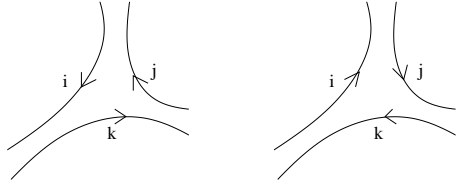}
\end{center}
\caption{ \footnotesize A-type Junction and Anti-Junction}
\label{junction}
\end{figure}

  Now there are 1/4 BPS waves on 1/2 BPS selfdual strings. As there are left and right moving waves for a given selfdual strings, there are $2\times N(N-1)/2=N(N-1)$ such objects. The 1/4 BPS junctions can exist for any three choice of M5 branes, or any three roots $\a,\b,\g$ such that their sum vanishes. One way to represent such roots is $\a=e_i-e_j, \b=e_j-e_k, \g=e_k-e_i$. The junction and its anti-junction for such three roots is shown in Figure~\ref{junction}.  The total number of junctions and anti-junctions would be $2\times N(N-1)(N-2)/6=N(N-1)(N-2)/3$.
The total number of 1/4 BPS objects  in the   6d (2,0) theory of the $A_{N-1}=SU(N)$ gauge  would be then
\be N(N-1)+ \frac13 N(N-1)(N-2)= \frac13 N(N^2-1) =\frac13 c_{A_{N-1}} . \ee
This matches one-third of the anomaly coefficient $c_{A_{N-1}}$. 
We would readily admit that this counting is naive at best as we have ignored the spin and other structures of these object. Also we have ignored additional degeneracy for the wave structure.

%
%
%
%

\begin{figure}[h]
\begin{center}
\leavevmode
\epsfxsize 9 cm
\epsffile{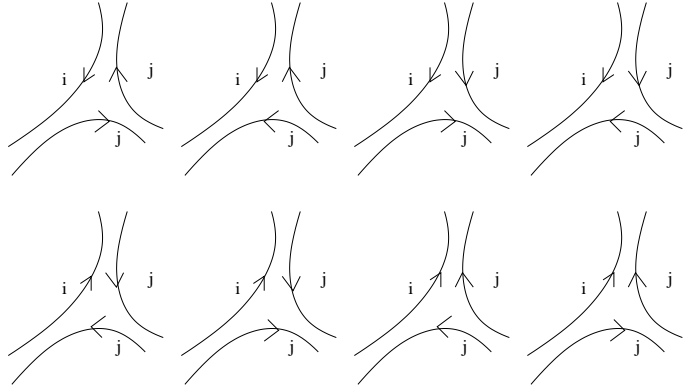}
\end{center}
\caption{\footnotesize D-type Junctions and Anti-Junctions}
\label{djunction}
\end{figure}

The counting of  1/4 BPS objects in the (2,0) theory of $D_N=SO(2N)$ gauge group goes similarly. The root vectors $\pm (e_i-e_j)$ or $\pm (e_i+e_j)$ where $i\neq j$ and $i,j=1,2,\cdots N$. There would $N$ 1/2 BPS massless particles and $ N(N-1)$ 1/2 BPS selfdual strings.   Now the number of the 1/4 BPS waves on   selfdual strings would be $2N(N-1)$. 
The counting of junctions and anti-junctions is a bit more complicated. At the junction, the sum of the root should vanish. The Figure~\ref{djunction} shows the types of junctions and anti-junctions. Thus there are eight types for the selection of $i\neq j\neq k\neq i$, leading to the total number of 1/4 BPS junctions to be
$8\times N(N-1)(N-2)/6=4N(N-1)(N-2)/3$.
The total number of the 1/4 BPS objects  in the 6d (2,0) theory of  $D_N=SO(2N)$ gauge algebra    in the
 generic Coulomb phase would be then
 \be 2N(N-1)+ \frac43 N(N-1)(N-2) = \frac23 N(N-1)(2N-1) =\frac13 c_{D_N}.  \ee
Again it matches one-third of the anomaly coefficient.

\begin{figure}[h]
\begin{center}
\leavevmode
\epsfxsize 12 cm
\epsffile{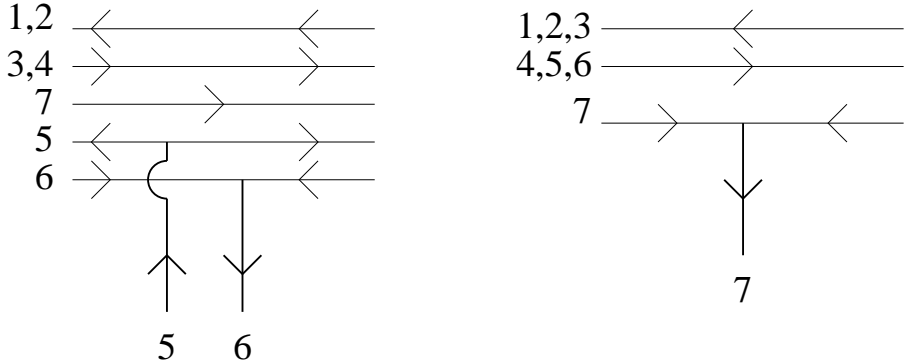}
\end{center}
\caption{Two Examples of E6  Junctions  }
\label{e6junction}
\end{figure}

The root diagram of $E_6$ is made of the root of $A_5=SU(6)$ and also $\pm \sqrt{2}e_7$ and
$\frac12 (\pm e_1\pm e_2\pm e_3\pm e_4\pm e_5\pm e_6)\pm e_7/\sqrt{2} $ with the number of plus sign for $e_1\cdots e_6$ being three.  Note that $d_{E_6}=78$ and $h_{E_6}=12$ and so $c_{E_6}/3=312$.   The number of the 1/4 BPS objects of $SU(6)$ is 70. The number of the additional 1/4 BPS wave on monopole is 42.
The number of additional 1/4 BPS junctions with one end being $e_7$ is 20. Finally the number of additional 1/4 BPS junctions with one end of type $e_i-e_j$ is 180.
Total 1/4 BPS object is then
\be 70+ 42+ 20+180=312 =\frac13 c_{E_6}. \ee
Figure~\ref{e6junction} shows   two types of additional junctions besides those from $A_5$. Note that the sum of the roots for these junctions vanishes. 

\begin{figure}[h]
\begin{center}
\leavevmode
\epsfxsize 5 cm
\epsffile{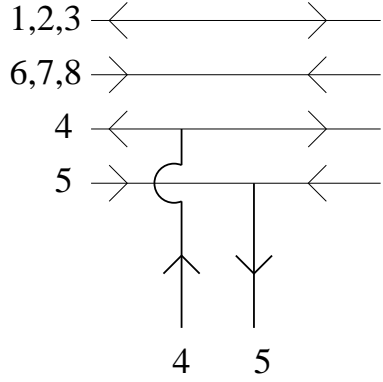}
\end{center}
\caption{An Example  of E7  Junctions  }
\label{e7junction}
\end{figure}

The root diagram of $E_7$ is made of the roots of $A_7=SU(8)$ and the roots $(\pm e_1\pm e_2\cdots\pm e_8)/2$ with four plus and  four minus signs. The number of 1/4 BPS objects from $A_7=SU(8)$ is 168. The number of additional 1/4 BPS waves on the string is 70. The number of additional 1/4 BPS junction is $8*7/2*2*1/2*6*5*4/6=560$. The total number of 1/4 BPS objects are
\be 168+ 70+560= 798 =\frac13 c_{E_7}. \ee
Figure~\ref{e7junction} shows an example of  additional junctions of $E_7$ case besides that from $A_7$. 

\begin{figure}[h]
\begin{center}
\leavevmode
\epsfxsize 12 cm
\epsffile{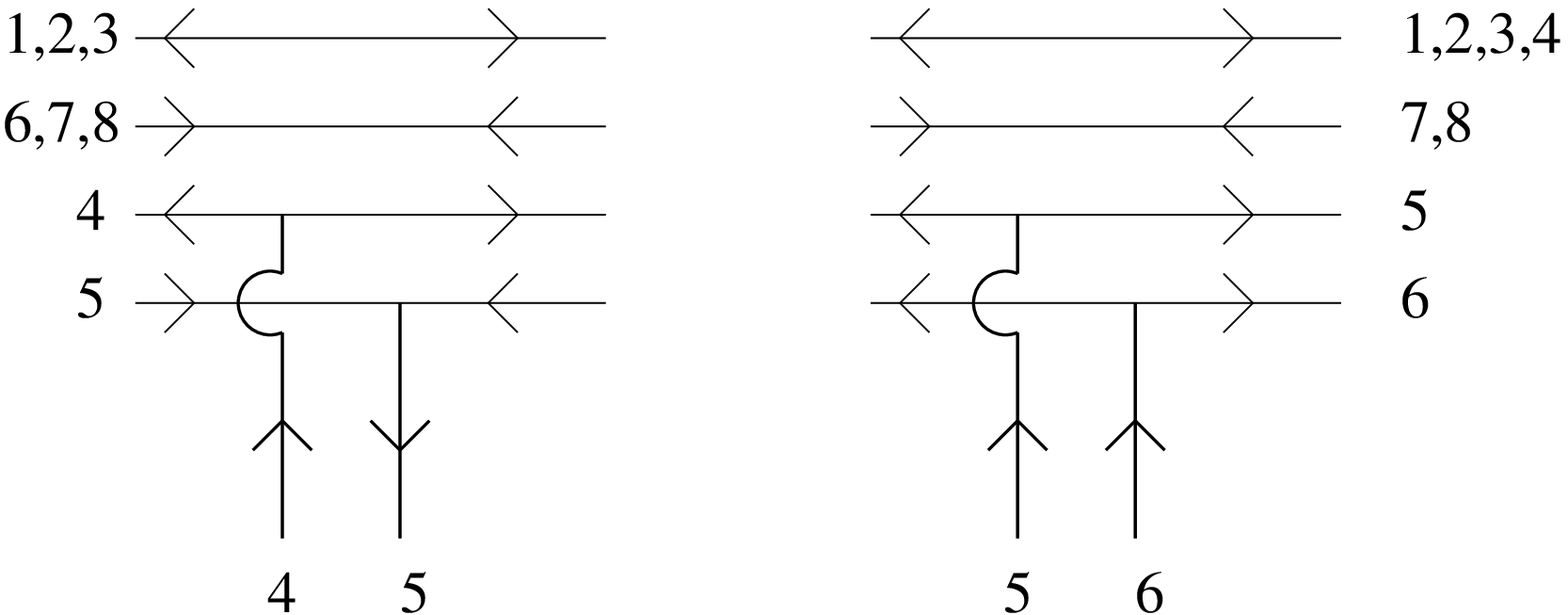}
\end{center}
\caption{Two Examples  of E8  Junctions  }
\label{e8junction}
\end{figure}

The root system of $E_8$ is made of the roots $\pm e_i \pm e_j$ $(i\neq j, 1,2\cdots 8)$ of $D_8=SO(16)$ and $\frac12(\pm e_1\pm e_2\cdots \pm e_8)$ with the product of the sign is plus one. The number of 1/4 BPS objects from $D_8=SO(16)$ theory is 560. The number of 1/4 BPS waves on additional self-dual strings is 128. The   number of additional junction is $8*7/2*2^8/2/2=1792$. Thus, the total number of 14 BPS objects for $E_8$ case is
\be 
560+128+1792=2480 = \frac13 c_{E_8}, 
\ee
which is exactly the number we expect.
Figure~\ref{e8junction} shows   two examples of additional junctions besides that from $A_7$. 

So far our counting of 1/4 BPS objects has been somewhat naive as we did not include the spin degrees of freedom. For example magnetic monopole strings would have the $\CN=(4,4)$ world sheet supersymmetries on the string. The 1/2 BPS selfdual strings would have the $\CN=(4,4)$ world sheet supersymmetries with torsion so that the left and right complex structure being different for two parallel identical strings. 1/4 BPS junctions would have more complicated spin structure, which is beyond the scope of the current work. There is a classification for the representation of this superalgebra in terms of superfields \cite{Ferrara:2000xg}.  This would be a starting point and we hope to return to this issue in future. 

We would like to add some further comments on  the mathematical structure of our observation.   By definition,   Coxeter number $h_G$ is the number of roots $d_G-r_G$ divided by the rank $r_G$, and so $d_G=(h_G+1)r_G$. Note that Coxeter number and dual Coxeter number coincide for simple-laced group. Our relation is then
\be \frac13 c_G =   h_G r_G + \frac13 r_Gh_G(h_G-2) \ee
The number of independent BPS junctions and anti-junctions is   twice the number of $SU(3)$ embeddings to the simple-laced group and is given by the last number~\cite{mathoverflow}. We make a further observation in relation to    the central charge (\ref{toda}) of the Toda models for simple   laced group\cite{Hollowood:1989} where it appeared via the Freudenthal and de Vries' strange formula, 
\be \frac13 c_A =\frac13 h_G^V d_G = 4{\boldsymbol{\rho} }^2 
\ee
where the Weyl vector  
\be   \boldsymbol{\rho}  =\frac12 \sum_{\a>0}\a \ee
is the sum over the positive root with the convention that the length square of a long root is two. Thus we get
\be \frac13 c_A = \sum_{\a>0} \a^2 + 2\sum_{ \a,\b >0 ,\a\neq \b  } \a\cdot\b  
\ee
Note that the first term counts the number of roots as $\a^2=2$ and the second term counts twice the number of $SU(3)$ embedding, or the number of junctions and anti-junctions.
 
\section{Concluding Remarks}

Let us conclude with some remarks. We have identified all 1/4 BPS objects and anti-objects in the 6d (2,0) superconformal theories  in the generic Coulomb phase. These 1/4 BPS objects consist of  waves on selfdual  strings and   junctions of selfdual strings. The total number of 1/4 BPS objects and anti-objects is exactly one third of the anomaly coefficient $c_G$ for all (2,0) theories.   These junctions would also contribute to the low energy effective Lagrangian for the abelian degrees of freedom. For $A_N, D_N$ for large $N$ case, $c_G\sim N^3$. This suggests that these 1/4 BPS objects may be  fundamental ones in the (2,0) theories at least in the Coulomb phase, and may play a key role in our understanding of the (2,0) theories even in the symmetric or conformal phase. For example, after local heading of M5 branes in Coulomb phase there could be `generalized Hagadron phase transition' which not only release selfdual string loops  but also junctions and anti-junction nets or more complicated webs of strings. 
While the numbers $c_G/3$ for AD type theories coincide  with the dimensions of some representations, it is not the case for E type theory. This implies the number $c_G/3$ cannot be represented as some objects of irreducible representation of the group $G$ in general.
Further studies may be needed to  relate these 1/4 BPS objects to the anomaly calculation.

\section*{Acknowledgement}

We are grateful to Sungjay Lee and Yuji Tachikawa   for helpful discussions. KL is supported in part by NRF-2005-0049409 through  CQUeST,
and the National Research Foundation of Korea Grants NRF-2009-0084601 and NRF-2006-0093850.
SB wants to thank KIAS for the hospitality in March 2011 when part of this work was done.

\end{document}